\documentclass[12pt, english]{article}
\usepackage{babel}

\begin{document}

\title{The impact hazard from small asteroids: current problems and open
questions}

\author{Luigi Foschini\\ 
{\small Istituto TeSRE - CNR, Via Gobetti 101, I-40129 Bologna, Italy}\\
{\small Email: foschini@tesre.bo.cnr.it}}

\date{}
\maketitle

\begin{abstract}
The current philosophy of impact hazard considers the
danger from small asteroids negligible. However, several facts claim for a
revision of this philosophy. In this paper, some of these facts are reviewed and
discussed. It is worth noting that while the impact frequency of
Tunguska--like objects seems to be higher than previously estimated,
the atmospheric fragmentation is more efficient than commonly thought.
Indeed, data recorded from airbursts show that small asteroids breakup at
dynamical pressures lower than their mechanical strength. This means that
theoretical models are inconsistent with observations and new models
and data are required in order to understand the phenomena. 
\end{abstract}

\section{Introduction}
The interest in the impact of interplanetary bodies with planets,
particularly with Earth, has been increased significantly during the last few
years because of several events such as the fall of the
D/Shoemaker--Levy~9 into Jupiter's atmosphere.

Particular attention was given to the detection of kilometre--sized
objects, which pose a severe threat to the Earth. In recent years,
this has been emphasized by several authors with
differing points of view (e.g. Aduskhin and Nemtchinov 1994,
Chapman and Morrison 1994, Toon et al.  1997).  The reason is
quite simple, as written by Clark Chapman (1996): the impact of such
an object has non--zero probability of creating a global ecological
catastrophe within our lifetime.

Larger objects (tens of kilometres) can cause an extinction
level event.  The consequent ``asteroidal winter'', deriving from a 
strong injection of dust in the atmosphere, is quite similar to the 
nuclear winter, radioactive consequences apart.  It would cause
the onset of environmental conditions whose main features are: a very 
long period of darkness and reduced global temperature,  something
similar to the polar winter on a world wide scale (Cockell
and Stokes 1999).

Even though I understand and respect these opinions, I think that we
cannot neglect small bodies at all. There are two main reasons: first, the
fragmentation of asteroids in the Earth's atmosphere is not well known.
Observations of small asteroids (up to tens of metres) show that the
fragmentation occurs when the dynamical pressure is lower than the mechanical
strength, and there is no reason to suppose that larger bodies behave
differently. Therefore, airburst can give us data to test theories for
fragmentation, which are also valid for larger bodies.

The second reason is that, although the damage caused by Tunguska--like
can be defined as ``local'', it is not negligible. Specifically, there are
several scientists, such as J. Lewis, M. Paine, S.P. Worden and B.J. Peiser
(see debates in the Cambridge Conference Net), suggesting that small 
asteroids might be even more dangerous than larger bodies.

Moreover, David Jewitt (2000), after the paper by Rabinowitz et al. (2000)
where authors strongly reduced the number of NEO larger than 1~km, suggested
that it is time to set up a more ambitious NEO survey, including small objects.

The present paper does not present any new theory or observation, but it review
some points that are not present in previous analyses and studies. The purpose of
this paper is to strengthen studies on small objects simply because 
our knowledge is very poor. 
The paper is divided into two parts: in the Section 2, I add some notes to the
debate on the danger from small asteroids. In the Sects. 3 and 4, I present
the evidence that the fragmentation of small asteroids in the Earth's atmosphere
is still an open problem.

\section{Tunguska--like events}
Small objects, of the order of tens or hundreds of metres, can cause 
severe local damages.  The best known event of this kind is the
Tunguska event of 30 June 1908, which resulted in the devastation of
an area of $2150\pm 25$~km$^2$ and the destruction of more than 60 
million trees (for a review, see Vasilyev 1998). Still today there
is a wide debate all over the world about the nature of the cosmic 
body which caused that disaster. Just last July an Italian
scientific expedition, \emph{Tunguska99}, went to Siberia to collect
data and samples (Longo et al. 1999).

Chapman and Morrison (1994) considered Tunguska--like events as a 
negligible threat. They could be right, considering the substantial uncertainties
in these studies, but they underestimate some values. Although
they proposed data with large error bars, the question is: where do we
have to center these bars?
 
Let us analyse the assumptions of Chapman and Morrison: 
first of all, they consider that the area destroyed in Tunguska
(i.e. the area where the shock wave was sufficient to fell trees) was
about 1000~km$^2$. This value is somehow larger than the area where the 
peak overpressure reached the value of 4~psi (27560~Pa), sufficient to
destroy normal buildings (according to the formula quoted by Chapman 
and Morrison the area of 4~psi is about 740~km$^2$, using a yield
of 20~Mton).  

There are two main objections to this hypotesis: firstly, the \emph{measured}
value of the area with fallen trees is more than double (see above;
Vasilyev 1998). In addition to this, it is worth noting that an 
overpressure of 2~psi produces wind of 30~m/s, which is sufficient
to cause severe damages to wood structures.  In addition to this, debris 
flying at such speed is a threat to life (Toon et al.  1997).

Therefore, a reasonable value of human beings \emph{risking death} during a
Tunguska--like event is $10^{4}$, rather than $7\times 10^{3}$ as indicated 
by Chapman and Morrison.  The above value has been calculated by 
using the formula in Adushkin and Nemtchinov (1994) and assuming an explosion 
energy of 12.5 Mton (Ben--Menahem 1975).

Chapman and Morrison (1994) correctly note that there is a much greater
probability that such an event might occur in a uninhabitated part
of the world.  On the other hand, in the unlikely event of it occurring in a
populated city, it would cause a great disaster.
For example, in Rome which has a population density of about 2000 people per
square kilometre the number of human beings at risk would be more than 2 millions.

It is also necessary to evaluate the impact frequency
of Tunguska--like events.  Chapman and Morrison consider a time
interval of 250 yr, but several other studies and episodes suggested a 
lower value.  Farinella and Menichella (1998) studied the 
interplanetary dynamics of Tunguska--sized bodies by means of a 
numerical model and they found that the impact frequency is 1 per 100 
yr.  However, in that study, the authors did not take into account the
Yarkovsky effect (see Farinella and Vokrouhlick\`{y}, 1999, and 
references therein), that can slightly increase the delivery of NEO 
(Near Earth Objects) toward the Earth.

There are also ground--based and space--based observations that 
support these conclusions, even though the frequency range can vary 
greatly.  For a 1~Mton explosion, the impact frequency can be once
in 17 (ReVelle 1997) or 40 yr (Nemtchinov et al.  1997b), that implies 
a Tunguska event (12.5~Mton) once in 100 or 366 yr.  If we consider an 
energy of 10~Mton, as calculated by Hunt et al.  (1960), we obtain a 
value of the impact frequency of respectively 88 or 302 yr.  In
addition to this, Steel (1995) reportes two other Tunguska--like event
in South America in 1930 and 1935: this strengthens the impact
frequency value of one per 100 yr (or less).

Now, if we consider a typical time interval of one Tunguska--like 
impact per 100 yr and $10^{4}$ deaths per impact, we obtain 100 death
per years throughout the world; this value is no longer negligible in
the Chapman and Morrison's scale (1994).

\emph{On the other hand, we would stress the great
uncertainty of these values, which are mainly due to the use of empirical
relations with scarce data.} We are aware that the threat 
posed by kilometre and multikilometre objects is more dangerous and 
therefore we must study these objects and methods to avoid a global 
catastrophe.  However, the few points raised in this paper
suggest that we must \emph{also} study Tunguska--like events. In 
addition to this, it is worth noting that studies about the impact 
hazard are often based on models of cosmic bodies fragmentation in 
the Earth's atmosphere. These models assume that the fragmentation 
begins when the dynamical pressure in the stagnation point is equal 
to the mechanical strength of the body. However, as we shall see, this 
does not occur.

\section{The failure of current theories}
The calculations of the impact hazard are strongly related to
available numerical models for the fragmentation of asteroids/comets 
in the Earth's atmosphere.  Present models consider that
fragmentation begins when the dynamical pressure in front of the cosmic
body is equal to the material mechanical strength.  However,
observations of very bright bolides proves that large meteoroids
or small asteroids breakup at dynamical pressures lower than their 
mechanical strength.  Today there is still no explanation
for this conundrum. This is of paramount 
importance, because it allow us to know whether or not an asteroid might
reach the Earth's surface.  In addition to this, the atmospheric breakup also
effects the crater field formation (Passey and Melosh 1980) or on
the area devastated by the airblast.  Therefore, it allows us to
establish a reliable criterium to assess the impact hazard.  All 
studies shown above are based on models where fragmentation begin
when the dynamical pressure is equal to the mechanical strength of the 
asteroid.  But, as we shall see, observations indicate that this is 
not true.

The interaction of a cosmic body in the Earth's atmosphere can be 
divided into two parts, according to the body dimensions.  For 
millimetre to metre sized bodies (meteoroids), the most useful 
theoretical model is the gross--fragmentation model developed by 
Ceplecha et al.  (1993) and Ceplecha (1999).  In this model, there are 
two basic fragmentation phenomena: \emph{continuous
fragmentation}, which is the main process of the meteoroid ablation, 
and \emph{sudden fragmentation} or the discrete fragmentation at a
certain point.

For small asteroids another model is used, where the ablation
is contained in the form of explosive fragmentation, while at high
atmospheric heights it is considered negligible.  Several models have been
developed: Baldwin and Shaeffer (1971), Grigoryan (1979), Chyba et al.
(1993), Hills and Goda (1993), Lyne et al.  (1996).  A comparative 
study on models by Grigoryan, Hills and Goda, and Chyba--Thomas-Zahnle 
was carried out by Bronshten (1995).  He notes that the model proposed
by Chyba et al.  does not take into account fragmentation:
therefore, the destruction heights are overestimated (about 
10--12~km).  Bronshten also concludes that the Grigoryan and
Hills--Goda's models are equivalent.

There is also a class of numerical models, called ``hydrocodes''
(e.g., CTH, SPH), which were used particularly for the recent impact 
of Shoemaker--Levy~9 with Jupiter.  Specifically, Crawford (1997) uses 
CTH to simulate the impact, while M. Warren, J. Salmon, M. Davies and P. Goda 
used SPH. The latter was only published on the internet and is no longer
available. 

Despite the particular features of each model,
fragmentation is always considered to start when the dynamical pressure $p_0$ 
in the front of the meteoroid (stagnation point) exceeds the 
mechanical strength $S$ of the body.

Although direct observations for asteroid impact are not available, it is 
possible to compare these models with observations of bodies with 
dimensions of several metres or tens of metres.  Indeed, in this 
range, the gross--fragmentation model overlaps the explosive 
fragmentation models.  As underlined several times by Ceplecha (1994, 
1995, 1996b), observations clearly show that meteoroids breakup at 
dynamical pressures lower (10 times and more) than their mechanical 
strength.  These data are obtained from photographic observation of
meteors and the application of the gross--fragmentation model, that 
can be very precise.  According to Ceplecha et al.  (1993)
it is possible to distinguish five strength categories with an average 
dynamical pressure of fragmentation (Tab.~\ref{category}).

\begin{table}[h]
	\centering
	\caption{Meteoroid strength category.  After Ceplecha et al.
	(1993)}
	\begin{tabular}{ccc}
	\hline
	Category & Range of $p_{\mathrm{fr}}$ [MPa] & Average 
	$p_{\mathrm{fr}}$ [MPa]\\
	\hline
	a & $p<0.14$ & $0.08$\\
	b & $0.14 \leq p < 0.39$ & $0.25$\\
	c & $0.39 \leq p < 0.67$ & $0.53$\\
	d & $0.67 \leq p < 0.97$ & $0.80$\\
	e & $0.97 \leq p < 1.2$ & $1.10$\\
	\hline
	\end{tabular}
	\label{category}
\end{table}

For continuous fragmentation the results obtained also indicate that
the maximum dynamical pressure is below 1.2~MPa, but five
exceptions were found: 4 bolides reached 1.5~MPa and one survived up to
5~MPa (Ceplecha et al.  1993).

It is also very important to relate the ablation coefficient $\sigma$ 
with the fragmentation pressure $p_{\mathrm{fr}}$, in order to find
a relationship between the meteoroid composition and its resistance to
the air flow.  To our knowledge, a detailed statistical
analysis on this subject does not exist, but in the paper by Ceplecha 
et al.  (1993) we can find a plot made by considering data on 30
bolides (we refer to Fig.~12 in that paper). We note that stony 
bodies (type I) have a wide range of $p_{\mathrm{fr}}$ values.
In the case of weak bodies, we can see that there is
only one cometary bolide (type IIIA), but this is due to two factors:
firstly, cometary bodies undergo continuous fragmentation, rather than a
discrete breakup at certain points. Therefore, it is incorrect to
speak about fragmentation pressure; we should use the maximum
tolerable pressure. The second reason is that there is a selection 
effect. Indeed, from statistical studies, Ceplecha et al. (1997) 
found that a large part of bodies in the size range from 2 to 15~m 
are weak cometary bodies. 

However, a recent paper has shown that statistics from physical 
properties can lead to different results when compared with statistics 
from orbital evolution (Foschini et al.  2000). To be more precise,
physical parameters prove that, as indicated above, a large
part of small near Earth objects are weak cometary bodies, whilst,
the analysis of orbital evolution proves a strong asteroidal
component.

The reason for the presence of cosmic bodies with very low
fragmentation pressure can be explained by the assumption that
additional flaws and cracks may be created by collisions in space, 
even though they do not completely destroy the cosmic body (Baldwin and 
Shaeffer 1971).  Other explanations could be that the asteroid was not
homogeneous (see the referee's comment in Ceplecha et al. 1996) or it 
had internal voids (Foschini 1998).

Almost all models described deal with the
motion of a cosmic body in the Earth's atmosphere. However, it is
worth noting that we cannot observe directly the cosmic body: we can
only see the light emitted during the atmospheric entry. Therefore, we have
to introduce in equations several coefficient that cannot be derived
from direct observations.

If we turn our attention to the hypersonic flow around the body, we
could have data from direct observations.  Among models discussed
above, only Nemtchinov et al.  (1997a, b) tried to investigate the
hypersonic flow around the asteroid with a numerical model.
Foschini (1999) investigated the analytic approach: indeed, although
the details of an hypersonic flow are very difficult to calculate and
there is need of numerical models, the pressure distribution can be
evaluated with reasonable precision by means of approximate methods.
In the limit of a strong shock ($M>>1$) several equations tend
to asymptotic values and calculations become easier. The application of this
technique to a particular episode, such as the Tunguska event, gave reasonable
values (Foschini 1999). However, although first results are encouraging, further work is
necessary before having a complete and detailed theory.

\section{Special cases}
In addition to data published in the paper by Ceplecha et al.  (1993) 
and Ceplecha (1994) we consider some specific cases of bright 
bolides.  We provide here a short description and we refer for details 
to the papers quoted.

\begin{table*}
\centering
\caption{Special episodes.}
\begin{tabular}{lrrr}
\hline
Name & Date & max $p_{\mathrm{fr}}$ [MPa] & $S$ [MPa]\\
\hline
P\v{r}\`{\i}bram & Apr 7, 1959 & 9.2 & 50\\
Lost City & Jan 3, 1970 & 1.5 & 50\\
\v{S}umava & Dec 4, 1974 & 0.14 & 1\\
Innisfree & Feb 6, 1977 & 1.8 & 10\\
Space based obs.  & Apr 15, 1988 & 2.0 & 50\\
Space based obs.  & Oct 1, 1990 & 1.5 & 50\\
Bene\v{s}ov & May 7, 1991 & 0.5 & 10\\
Peekskill & Oct 9, 1992 & 1.0 & 30\\
Marshall Isl.  & Feb 1, 1994 & 15 & 200\\
\hline
\end{tabular}
\label{special}
\end{table*}

The Lost City meteorite (January 3, 1970), a chondrite (H), was 
analysed by several authors (McCrosky et al.  1971, ReVelle 1979, 
Ceplecha 1996a).  The recent work by Ceplecha (1996a) is of particular 
interest, because by taking into account the meteoroid rotation, he 
succeeds in explaining the atmospheric motion without discrepancies.
Obviously, except the dynamical pressure, that in this episode
reaches the value of $p_{\mathrm{fr}}=1.5$~MPa, while the mechanical
strength of a stony body is about 50~MPa.

In the work by ReVelle (1979), it is also possible to find useful data
for two other episodes: P\v{r}\`{\i}bram (April 7, 1959) and Innisfree 
(February 6, 1977).  In both episodes a meteorite was recovered: respectively
ordinary chondrite and L chondrite.  Values for
$p_{\mathrm{fr}}$ of 9.2~MPa and 1.8~MPa respectively were obtained in 
this work.

The \v{S}umava bolide (December 4, 1974) reached $-21.5$ absolute 
visual magnitude and was produced by a cometary body.  It exhibited 
several flares during continuous fragmentation, ending at a height of
about 60~km.  The maximum dynamical pressure was in the range
$0.025-0.14$~MPa, much lower than the mechanical strength of a
cometary body, i.e.  1~MPa (Borovi\v{c}ka and Spurn\'y 1996).

The Bene\v{s}ov bolide (May 7, 1991) was very atypical and was analysed
in detail by Borovi\v{c}ka and Spurn\'y (1996) and Borovi\v{c}ka et 
al.~(1998a, b).  From these studies, results show that it was very
probably a stony object which underwent a first fragmentation at high
altitudes ($50-60$~km) at dynamical pressures of about $0.1-0.5$~MPa.  
However, some compact fragments were disrupted at pressures of 9~MPa 
(24~km of height).  

The fall of the Peekskill meteorite (October 9, 1992) was the first of 
such events to be recorded by a video camera (Ceplecha et al.  1996).  
The fireball was brighter than the full moon and 12.4~kg of ordinary 
chondrite (H6 monomict breccia) were recovered.  Tha availability of a 
video recording allows us to compute, with relative precision, the evolution
of the meteoroid speed and, therefore, the dynamical pressure.  It
was discovered that the maximum value of $p_{\mathrm{fr}}$ was about
$0.7-1.0$~MPa, while the meteorite has an estimated strength close to 
30~MPa.

In recent years, space--based infrared sensors detected several
bolides all around the world.  Nemtchinov et al.  (1997) investigated 
these events by using a radiative--hydrodynamical numerical code.  
They simulated three bright bolides (April 15, 1988; October 1, 1990; 
February 1, 1994) and they obtained respectively these results: stony 
meteoroid, $p_{\mathrm{fr}}=1.6-2.0$~MPa; stony meteoroid, 
$p_{\mathrm{fr}}=1.5$~MPa; iron meteoroid, 
$p_{\mathrm{fr}}=10-15$~MPa.  Concerning the latter, Tagliaferri
et al.  (1995) reached a slightly different conclusion: stony 
meteoroid, $p_{\mathrm{fr}}=9$~MPa.

The condition that fragmentation starts when the dynamical
pressure reaches the mechanical strength of the meteoroid was imposed 
by Baldwin and Shaeffer (1971), but it is worth noting that this is a
hypotesis.  Now we have sufficient, though incomplete, data to
claim that this hypotesis has no physical ground and we have to find
new conditions for fragmentation.

\section{Conclusion}
Only in recent decades, and particularly in recent years, the impact
hazard has attracted the attention of more and more scientists. Evaluation
of impact frequencies and damages are made by means of empirical or
semiempirical formulas. However, we are faced with scarce, and often
contradictory data. For example, Chapman and Morrison (1994) considered an
impact frequency of one Tunguska--like event every 250 yr by using data from
lunar craters, ReVelle obtains a higher frequency for the same kind of objects
(1 per 100 yr) by considering data from airbursts.

The main problem is the fragmentation mechanism, that is still
unclear. From observations, it results that fragmentation occurs when the
dynamical pressure is lower than the mechanical strength. We do not know whether
this is due to any special feature in the hypersonic flow around the body or to
any particular matter in the body. Today all that we can say is that current models
of fragmentation of small asteroids in the Earth's atmosphere
\emph{are not consistent} with observations. We require more data and
theories to understand the matter better. Airbursts can give us useful data to
test theories.

\section{Acknowledgements}
Part of this work was already presented at the IMPACT Workshop (1999). 
I wish to thank the International Astronomial Union for a grant that 
allowed me to attend the IMPACT Workshop in Torino.  
Some ideas exposed here rose thanks to 
discussions with Zdenek Ceplecha during a visit in the Ond\v{r}ejov 
Astronomical Observatory: I wish to thank Z.~Ceplecha, his wife 
Hana, and meteor scientists at the observatory for their kind hospitality.

\end{document}